\documentclass[prb,twocolumn,showpacs,amsmath,amssymb,floatfix]{revtex4}

\usepackage[]{graphicx}		
\usepackage{bm}				
\usepackage{tabularx}

\newcommand{\erfc}{\ensuremath{\textrm{erfc}}}
\newcommand{\sech}{\ensuremath{\textrm{sech}}}

\begin{document}

\title{Influence of magnetic viscosity on domain wall dynamics under spin-polarized currents}
\author{Joo-Von Kim}
\email{joo-von.kim@u-psud.fr}
\author{C. Burrowes}
\affiliation{Institut d'Electronique Fondamentale, CNRS, UMR 8622, 91405 Orsay, France} 
\affiliation{Universit\'{e} Paris-Sud, UMR 8622, 91405 Orsay, France} 
\date{\today}

\begin{abstract}
We present a theoretical study of the influence of magnetic viscosity on current-driven domain wall dynamics. In particular we examine how domain wall depinning transitions, driven by thermal activation, are influenced by the adiabatic and nonadiabatic spin-torques. We find the Arrhenius law that describes the transition rate for activation over a single energy barrier remains applicable under currents but with a current-dependent barrier height. We show that the effective energy barrier is dominated by a linear current dependence under usual experimental conditions, with a variation that depends only on the nonadiabatic spin torque coefficient $\beta$.
\end{abstract}


\maketitle

\section{Introduction}

Spin-polarized currents flowing through ferromagnetic media exert torques on local magnetic moments in regions where a spatial magnetization gradient occurs.~\cite{Berger:JAP:1984} This interaction originates from an exchange interaction between the charge carrier spins and local magnetization, and has been observed in systems based on transition ferromagnet metals~\cite{Freitas:JAP:1985,Klaui:APL:2003,Grollier:APL:2003} and dilute magnetic semiconductors~\cite{Yamanouchi:Nature:2004}  alike. In the context of domain walls, such spin-torques can be sufficiently large to induce wall motion even in the absence of magnetic fields.~\cite{Vernier:EPL:2004} Interest in current-driven domain wall dynamics stems from both fundamental considerations, whereby the problem involves reconciling complex spin-dependent transport processes with nonlinear magnetization dynamics, and potential applications for emerging spintronic technologies, such as novel logic circuits~\cite{Allwood:Science:2005} and nonvolatile memory devices.~\cite{Parkin:Science:2008,Hayashi:Science:2008}

The salient features of current-driven domain wall dynamics are captured in the modified Landau-Lifshitz equation of motion for the unit magnetization $\vec{m}$,~\cite{Zhang:PRL:2004,Thiaville:EPL:2005}
\begin{equation}
\frac{\partial \vec{m}}{\partial t} + \gamma_0 \vec{m} \times \vec{H}_{\rm eff} - \alpha \vec{m} \times \frac{\partial \vec{m}}{\partial t} = -(\vec{u} \cdot \nabla) \vec{m} + \beta \vec{m} \times (\vec{u} \cdot \nabla)\vec{m},
\label{eq:LLG}
\end{equation}
where the right-hand side describes the current-dependent terms. Here, $u$ represents an effective spin current drift velocity (see Section II for definition). The first term on the right hand side describes the adiabatic contribution, which arises from transport processes in which the charge carrier spins track the local magnetization variations as they traverse the wall. In contrast the second term, proportional to $\beta$, represents nonadiabatic effects. There remains much debate over the magnitude of the nonadiabatic torques,~\cite{Tatara:PRL:2004, Waintal:EPL:2004, Barnes:PRL:2005, Xiao:PRB:2006, Kohno:JPSJ:2006, Vanhaverbeke:PRB:2007, Thorwart:PRB:2007, Piechon:PRB:2007, Duine:PRB:2007, Tatara:JPSJ:2007, Tatara:JPSJ:2008, Lemaho:PRB:2009} whereby theoretical estimates invoking %
momentum transfer,~\cite{Tatara:PRL:2004, Tatara:JPSJ:2007, Tatara:JPSJ:2008}%
spin-mistracking,~\cite{Waintal:EPL:2004,Vanhaverbeke:PRB:2007} %
spin-flip scattering,~\cite{Kohno:JPSJ:2006, Duine:PRB:2007, Thorwart:PRB:2007, Tatara:JPSJ:2008} %
or even apparent nonadiabatic effects involving magnon emission,~\cite{Lemaho:PRB:2009} differ at least by an order of magnitude. The nonadiabatic term is important because it allows a domain wall to be driven into the viscous regime~\footnote{In field-driven dynamics, the viscous regime is characterized by a streaming wall motion for which the steady-state wall velocity is given by $v = \gamma \mu_0 H_a \lambda / \alpha$.}%
solely by applied currents with no current thresholds in perfect systems.~\cite{Thiaville:EPL:2005,Tatara:JPSJ:2006} This is in stark contrast to the effect of a pure adiabatic torque, whereby a current threshold governed by the hard-axis anisotropy needs to be surmounted before pure current-driven wall motion is possible.~\cite{Tatara:PRL:2004} Such dynamics can be readily deduced from the reduced equations of motion for a one-dimensional Bloch wall, which can be obtained from (1) by assuming a specified wall profile and integrating over the spatial coordinates,~\cite{Thiaville:EPL:2005}
\begin{subequations}
\begin{eqnarray}
\dot{X}_0/\lambda - \alpha \dot{\phi} &=& \gamma\mu_0 H_{\perp} \sin{\phi}\cos{\phi}
 + u/\lambda,  \\
\dot{\phi} +  \alpha \dot{X}_0/\lambda &=& \gamma\mu_0 H_a + \beta u/\lambda,
\end{eqnarray}
\end{subequations}
where $X_0$ denotes the wall position, $\phi$ describes a tilt angle that is associated with the conjugate momentum to $X_0$, $\lambda$ is the domain wall width, and $H_\perp$ is a transverse anisotropy field. From these equations it is immediately apparent that the nonadiabatic term plays a similar role to that of an applied magnetic field $H_a$, which has led to some attempts of quantifying $\beta$ experimentally by associating field-like variations, such as a reduction in the propagation field under currents, directly with the nonadiabatic torque. 

Despite the lack of consensus over the theoretical magnitude of the nonadiabatic torque, it is generally expected that large nonadiabatic effects should appear in systems with narrow domain walls. Good candidates to exhibit large nonadiabatic effects are therefore ferromagnetic alloy films with perpendicular anisotropy, such as Co/Pt and Co/Ni multilayers~\cite{Ravelosona:PRL:2005,Ravelosona:APL:2007,Burrowes:APL:2008} or FePt alloys,~\cite{Attane:PRL:2006} as wall widths in such materials can reach down to near-atomic dimensions of 1 nm. Furthermore, wall structures in these materials are expected to closely resemble Bloch domain walls, because the shape anisotropy associated with magnetization perpendicular to the film plane has the same symmetry as the intrinsic uniaxial anisotropy.

However, it is well established that wall motion in perpendicular anisotropy materials is largely dominated by intrinsic defects, to the extent that pinning fields can potentially be larger than the Walker breakdown field.~\cite{Metaxas:PRL:2007} For this reason, it is expected that thermal activation~\cite{Neel:1949,Brown:PR:1963} and magnetic viscosity~\cite{Street:1949} to be important for current-driven dynamics in such systems. In this context, the basis for the association of the nonadiabatic term with an effective field is unclear. While for the streaming dynamics the $\beta$ term enters the equations of motion like a field, a linear relationship between an applied magnetic field and an energy barrier, associated with pinning potentials due to defects, for example, does not necessarily exist in general.~\cite{Gaunt:JAP:1986} In this article we show that an Arrhenius law remains valid for describing transition rates $\tau^{-1}$ associated with thermally activated wall processes under spin-polarized currents, $\tau = \tau_{0} \exp(E_b/k_B T) $, but with a current-dependent effective energy barrier that depends only on nonadiabatic torques, $E_b = E_b(\beta I)$. Under usual experimental conditions, we show that the dominant contribution is a linear current term, which suggests that a direct experimental measure of $\beta$ can be obtained from a linear variation of the characteristic Arrhenius rate with current for domain wall depinning from a defect. As a corollary, the association of $\beta$ with an effective magnetic field is only possible if the energy barrier varies linearly with applied magnetic fields.

This article is organized as follows. In Section II, we present the one-dimensional Bloch wall model for wall dynamics. In Section III, we derive the Langevin and Fokker-Planck equations for the stochastic wall dynamics, and solve for the stationary probability distribution function. In Section IV, we calculate the current-dependent Arrhenius rate equation for thermal-activation of the domain wall over an energy barrier, and apply these results to depinning processes associated with point defects in Section V. We present some discussion and concluding remarks in Section VI.

\section{One-dimensional Bloch wall model}

We consider the effects of magnetic viscosity on the dynamics of a one-dimensional Bloch wall under spin-polarized currents. Bloch wall structures are expected to arise in magnetic films with perpendicular anisotropy, as the stray dipolar fields in this geometry act only to renormalize the unaxial perpendicular anisotropy. We will concern ourselves with the one-dimensional motion of a Bloch wall in a magnetic wire. Let $z$ represent the axis perpendicular to the film plane, which is collinear with the axis of uniaxial anisotropy $K_u$. Let $X$ represent the direction of the long axis of the wire, which corresponds to the direction in which the magnetization variation constituting the domain wall arises. It is assumed that the magnetization remains uniform along the $y$ and $z$ directions along the wire. Using spherical coordinates to represent the spatial magnetization distribution $\vec{m}(X) = (m_X,m_y,m_z) = (\cos\phi \sin\theta, \sin\phi \sin\theta, \cos\theta)$, where $[\theta,\phi] = [\theta(X),\phi(X)]$, the simplest magnetic Hamiltonian leading to a domain wall is
\begin{equation}
E_w = \int dV \left( A \, \frac{\partial m_i}{\partial X}\frac{\partial m_i}{\partial X} - K_u m_z^2 \right),
\label{eq:hamiltonian}
\end{equation}
where the first term represents the exchange interaction (with summation over repeated indices implied) with the exchange constant $A$, and the second term represents the uniaxial anisotropy. The static Bloch wall profile can be obtained using a variational procedure by minimizing the energy functional (\ref{eq:hamiltonian}) with respect to $[\theta,\phi]$; one readily finds $\theta(X) = 2 \tan^{-1} \exp[-(X-X_0)/\lambda]$, where the characteristic wall width is $\lambda = \sqrt{A/K_u}$. Note that in magnetic films with perpendicular anisotropy, the constant $K_u$ represents an effective anisotropy $K_u = K_{u,0} - \mu_0 M_s^2/2$ that represents the difference between the magnetocrystalline term $K_{u,0}$ and the demagnetizing fields due to the film geometry.

The salient features of the magnetization dynamics of a domain wall can be accounted for by the one-dimensional Bloch wall model. In this approach, the dynamics is simplified by integrating out the spatial degrees of freedom in the Landau-Lifshitz-Gilbert equation of motion (\ref{eq:LLG}). This is achieved by elevating the domain wall position $X_0(t)$ to a dynamical variable and by assuming that $\phi = \phi(t)$ is spatially uniform. Within this approximation, the wall dynamics is entirely parametrized in terms of the two wall coordinates $(X_0,\phi)$. In the absence of thermal fluctuations, the equations of motion are given by (including the adiabatic and nonadiabatic terms),~\cite{Thiaville:EPL:2005}
\begin{subequations}
\label{eq:EOM}
\begin{eqnarray}
\frac{1}{\lambda}\frac{\partial X_0}{\partial t} - \alpha \frac{\partial \phi}{\partial t} &=& \left(\frac{\gamma}{2 M_s S_c \lambda}\right)\frac{\partial E}{\partial \phi} + \frac{u}{\lambda},  \\
\frac{\partial \phi}{\partial t} +  \frac{\alpha}{\lambda}\frac{\partial X_0}{\partial t} &=& -\left(\frac{\gamma}{2 M_s S_c}\right)\frac{\partial E}{\partial X_0} + \frac{\beta u}{\lambda},
\end{eqnarray}
\end{subequations}
where $u = j P g \mu_B / 2 e M_s$ represents an effective spin current drift velocity, which accounts for the spin-transfer torque, $\gamma$ is the gyromagnetic constant, and $S_c$ is the cross-sectional area of the wire. $P$ is the spin polarization, $g$ is the gyromagnetic ratio, $\mu_B$ is the Bohr magneton, $e$ is the electron charge and $M_s$ is the saturation magnetization. The energy $E(X_0,\phi)$ represents the total magnetic energy of the domain wall, which includes the Zeeman energy from applied and stray dipole fields, and contributions from local field and anisotropy variations due to defects.

This set of coupled differential equations can be simplified using dimensional analysis by scaling out the characteristic length, time, and energy scales. The domain wall width $\lambda$ is the obvious characteristic length scale, which leads us to define the reduced spatial variable $x$
\begin{equation}
x = \frac{X_0}{\lambda}.
\end{equation}
A dimensionless time variable can be defined in terms of the characteristic frequency $\omega_{m} = \gamma \mu_0 M_s$,
\begin{equation}
\bar{t} = \omega_m t.
\end{equation}
In a similar way, we can define a dimensionless energy $\epsilon = E/E_0$, where the choice of
\begin{equation}
E_0 = \frac{2 M_s S_c \lambda \omega_m}{\gamma} = 2 \mu_0 M_s^2 (S_c \lambda) 
\end{equation}
allows the equations of motion to be simplified neatly. For a ferromagnetic material with $\mu_0 M_s = 0.75$ T and $\lambda = 10$ nm, patterned into a wire geometry with a cross-sectional area of $S_c = 100$ nm $\times 1$ nm, $E_0$ corresponds to a temperature of $6.5 \times 10^4$ K. In the same spirit, we will find it convenient to define a dimensionless temperature $\bar{T} = T/T_0$, where $T_0 = E_0/k_B$.

With these definitions and $\bar{u} = u / (\lambda \omega_m)$ being the dimensionless spin drift velocity, we obtain the dimensionless form of the equations of motion,
\begin{subequations}
\begin{eqnarray}
\frac{\partial x}{\partial \bar{t}} - \alpha \frac{\partial v}{\partial \bar{t}} &=& \frac{\partial \epsilon}{\partial v} + \bar{u},  \\
\frac{\partial v}{\partial \bar{t}} + \alpha \frac{\partial x}{\partial \bar{t}} &=& -\frac{\partial \epsilon}{\partial x} + \beta \bar{u},
\end{eqnarray}
\end{subequations}
where we have used the fact that the wall angle $\phi$ is linear in wall velocity at low velocities, which allows us to associate $\phi$ with a dimensionless velocity $v$, $v \equiv \phi$. This association makes the analogy between the 1D Bloch wall dynamics and the kinematics of a point particle more explicit. In this context, we note that the kinetic energy of the domain wall is associated with the excursion angle $\phi$ out of the wall plane thorugh the hard-axis anisotropy,
\begin{equation}
E_{\rm kin} = \frac{1}{2}\mu_0 M_s^2 \int dV \sin^2{\theta} \sin^2{\phi} = \mu_0 M_s^2 S_c \lambda \sin^2{\phi},
\end{equation}
which upon substituting the characteristic energy scale and $v$, and including a particle potential $U(x)$, leads to the reduced domain wall energy
\begin{equation}
\epsilon = \frac{1}{2} v^2 + U(x)
\end{equation}
which is valid at low velocities ($\sin^2\phi \approx \phi^2 = v^2$). The term $U(x)$ contains the Zeeman energy and any pinning potentials.

%
\section{Langevin equations and Fokker-Planck theory}
In order to describe processes involving thermal activation, we need to extend the theory to include thermal fluctuations. This can be achieved by adding stochastic forces $\eta$ to the equations of motion,~\cite{Duine:PRL:2007}
\begin{subequations}
\label{eq:langevin}
\begin{eqnarray}
\frac{\partial x}{\partial \bar{t}} - \alpha \frac{\partial v}{\partial \bar{t}} &=& \frac{\partial \epsilon}{\partial v} + \bar{u} + \eta_v(t), \\
\frac{\partial v}{\partial \bar{t}} + \alpha \frac{\partial x}{\partial \bar{t}} &=& -\frac{\partial \epsilon}{\partial x} + \beta \bar{u} + \eta_x(t),
\end{eqnarray}
\end{subequations}
where the $\eta$ terms represent white Gaussian processes with zero mean, $\langle \eta(t) \rangle = 0$, with two-time correlation functions of the form $ \langle \eta_\mu (t) \eta_\nu (t') \rangle = (2 \alpha \bar{T}) \delta_{\mu,\nu}\delta(t-t')$. This definition of the noise is consistent with the fluctuation-dissipation theorem for the closed thermodynamic system $\bar{u} = 0$.

Next, we seek to derive the Fokker-Planck equation that is associated with the coupled Langevin equations in Eq.~\ref{eq:langevin}. The Fokker-Planck equation describes the time evolution of the probability density $P(x,v,t)$ of finding the system in a state $(x,v)$ at time $t$. We proceed by first rewriting (\ref{eq:langevin}) to read
\begin{subequations}
\label{eq:langevin_red}
\begin{eqnarray}
(1+\alpha^2) \frac{\partial v}{\partial \bar{t}} &=& -\left( \frac{\partial \epsilon}{\partial x} + \alpha \frac{\partial \epsilon}{\partial v} \right) + \bar{u} (\beta - \alpha) \notag \\ 
&& + (\eta_x - \alpha \eta_v), \\
(1+\alpha^2) \frac{\partial x}{\partial \bar{t}} &=& \left( \frac{\partial \epsilon}{\partial v} - \alpha \frac{\partial \epsilon}{\partial x} \right) + \bar{u} (1+\alpha \beta) \notag \\
&& + (\alpha \eta_x + \eta_v).
\end{eqnarray}
\end{subequations}
Now, for a set of coupled nonlinear Langevin equations of the form,
\begin{equation}
\dot{x}_i = h_i(\{x_i\},t) + g_{ij} (\{x_i\},t) \eta_j(t),
\end{equation}
where $x_i$ represent the state variables and $\eta_i$ are the random fields, the corresponding Fokker-Planck equation for the probability density $P$ of finding the system in a state $\{ x_i \}$ at time $t$,
\begin{equation}
\frac{\partial P(\{x_i\},t)}{\partial t} = \hat{L}_{\rm FP} P(\{x_i\},t),
\end{equation}
with the Fokker-Planck operator
\begin{equation}
\hat{L}_{\rm FP} = -\frac{\partial}{\partial x_i} D_i(\{ x_i \},t) + \frac{\partial^2}{\partial x_i \partial x_j} D_{ij}(\{ x_i \},t),
\end{equation}
can be constructed from the drift coefficients
\begin{equation}
D_i(\{x_i \},t) = h_i(\{x_i\},t) + g_{kj}(\{x_i \},t) \frac{\partial}{\partial x_k} g_{ij}(\{x_i \},t)
\end{equation}
and diffusion tensors
\begin{equation}
D_{ij}(\{x_i \},t) = g_{ik}(\{x_i \},t)g_{jk}(\{x_i \},t),
\end{equation}
where summation over repeated indices is implied.~\cite{Risken:1989} By applying this prescription to the Langevin equations in (\ref{eq:langevin_red}), we find for the domain wall system
\begin{equation}
\hat{L}_{\rm FP} = \hat{L}_0 + \hat{L}_u,
\end{equation}
where
\begin{subequations}
\label{eq:FPop}
\begin{eqnarray}
\hat{L}_0 &=& -\frac{\partial}{\partial x}\left( \frac{\partial \epsilon}{\partial v} - \alpha \frac{\partial \epsilon}{\partial x}  \right) + \frac{\partial}{\partial v} \left( \frac{\partial \epsilon}{\partial x} + \alpha \frac{\partial \epsilon}{\partial v}  \right) \notag \\ && + \alpha \bar{T}\left( \frac{\partial^2}{\partial x^2} + \frac{\partial^2}{\partial v^2}  \right),  \\
\hat{L}_u &=& - \bar{u} \left[ (1+\alpha\beta) \frac{\partial}{\partial x} + (\beta-\alpha)\frac{\partial}{\partial v}  \right],
\end{eqnarray}
\end{subequations}
which agrees with the one-dimensional low-velocity limit of the operator found by Duine et al.~\cite{Duine:PRL:2007} Without loss of generality, we have assumed $\alpha^2 \ll 1$ to simplify the notation.

In the absence of currents, it can be easily verified that the Boltzmann distribution function satisfies the stationary Fokker-Planck equation,
\begin{equation}
\hat{L}_0 P_0(x,v) = 0,
\end{equation}
where
\begin{equation}
P_0(x,v) = Z^{-1} e^{-\epsilon/\tilde{T}},
\end{equation}
with $Z$ being a normalization constant. In the presence of currents, it is also possible to find the stationary solution to the full Fokker-Planck equation. From inspection of (\ref{eq:langevin}), we note that the adiabatic torque can be assimilated as a shift in the wall energy $\epsilon \rightarrow \epsilon + \bar{u}v$, while the nonadiabatic torque leads to the shift $\epsilon \rightarrow \epsilon - \bar{u}\beta x$, leading to the ansatz
\begin{equation}
P_0(x,v) = Z^{-1} \exp \left[ -\frac{\epsilon + \bar{u}(-\beta x + v)}{\tilde{T}} \right],
\label{eq:statdist}
\end{equation}
which is readily verified to be a solution to the stationary Fokker-Planck problem for $\bar{u} \neq 0$.

This distribution function shows that the Gaussian distribution over the particle ``velocity'' $v$ becomes centered around the value $v = -\bar{u}$ as a result of the adiabatic torque, $P_0(v) \propto \exp[\frac{1}{2}(v+\bar{u})^2]$. This would appear to be an inconsistent result, whereby the net domain wall flow takes place in a direction \emph{opposite} to the spin current. The origin of this apparent conundrum lies in the association of $\sin(\phi)$ with a wall velocity $v$. It is well established that while the adiabatic torque cannot drive the wall into a streaming motion at steady state (such as that achieved with a magnetic field or with a sufficiently large adiabatic torque~\cite{Thiaville:EPL:2005}), it does lead to a transient displacement of the wall with a finite value of $\phi$ at rest. The shift in the Gaussian distribution is a manifestation of this phenomenon, which restates that the tilt angle $\phi$ attains a finite value under adiabatic torques, but does not correspond directly to a finite streaming wall velocity.

\section{Arrhenius transition rate}
\begin{figure}
\includegraphics[width=5.5cm]{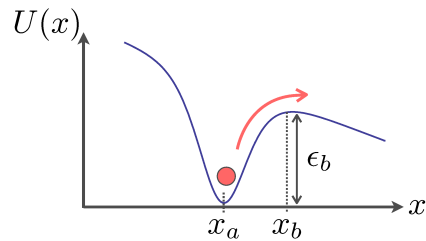}
\caption{\label{fig:potential_well}(Color online) Escape of a particle over an energy barrier by thermal activation. The barrier height is $\epsilon_b = U(x_b) - U(x_a) > 0$ and the characteristic potential well width is $\Delta x = x_b - x_a > 0$.}
\end{figure}

To calculate the thermally-activated depinning transition rate for a domain wall pinned at a defect, we follow the transition rate theory due to Kramers.~\cite{Kramers:1940,Hanggi:RMP:1990} This approach relies on drawing a strong analogy between the stochastic dynamics of the domain wall with the Brownian motion of a point particle in one-dimension. We consider a particle in a potential well as shown in Figure~\ref{fig:potential_well}, with an energy barrier given by $\epsilon_{b,0} = U(x_b) - U(x_a) > 0$.
The transition rate $\tau^{-1}$ of particles out of the potential well can be estimated by~\cite{Hanggi:RMP:1990}
\begin{equation}
\frac{1}{\tau} = \frac{j_b}{n_a},
\label{eq:tau}
\end{equation}
where $j_b$ is the particle current at the summit of the potential barrier $x = x_b$,
\begin{equation}
j_b = \int_{-\infty}^{\infty} dv \; J_x(x_b,v),
\end{equation}
with $J_x$ being the probability current density, and $n_a$ is the particle population in the well,
\begin{equation}
n_a = \int_{\rm well} dx \int_{-\infty}^{\infty} dv \; P_0(x,v).
\end{equation}

To compute the well population, it is convenient to use a quadratic expansion of the potential well around $x=x_a$,
\begin{equation}
U(x) \approx U(x_a) + \frac{1}{2} \omega_{a}^2 (x-x_a)^2,
\end{equation}
where $\omega_a^2>0$ is the square of a characteristic angular frequency that describes the curvature of the potential well at $x=x_a$. We assume that the particles injected into the well possess energies at least a few $k_B T$ below the barrier height and are rapidly thermalized before thermal activation processes take place, which allows us to use the steady-state distribution to describe the particle population at the bottom of the well. By using the stationary distribution function given in (\ref{eq:statdist}), we find
\begin{eqnarray}
n_a &=& \frac{1}{Z} \frac{2\pi \tilde{T}}{\omega_a} \\
&&\times \exp\left\{ \frac{1}{\bar{T}} \left[-U(x_a) +  \bar{u} \beta x_a + \frac{\bar{u}^2}{2} \left(1+ \frac{\beta^2}{\omega_a^2} \right) \right]  \right\}. \notag
\label{eq:na}
\end{eqnarray}

To describe the particle current at the barrier summit, we follow Kramers' approach by seeking a modified solution to the stationary Fokker-Planck equation of the form
\begin{equation}
P_0(x,v) = Z^{-1} \zeta(x,v) \exp \left[ -\frac{\epsilon + \bar{u}(-\beta x + v)}{\tilde{T}} \right],
\end{equation}
where the function $\zeta(x,v)$ accounts for sources, i.e. the injection of particles into the well at $x=x_a$, and sinks, whereby particles leaving the well $x>x_b$ are removed from the system. As before, we can approximate the potential energy at the barrier summit as
\begin{equation}
U(x) \approx U(x_b) - \frac{1}{2} \omega_b^2 (x-x_b)^2,
\end{equation}
where $\omega_b^2 > 0$. The limiting behavior required for $\zeta(x,v)$ is that it should be 1 in the well and tend toward zero for $x>x_b$. This behavior is only possible if $\zeta(x,v)$ depends on a linear combination of $x$ and $v$ only.~\cite{Kramers:1940} Let
\[
\xi = (x-x_b) + a (v+\bar{u}) + \frac{\beta \bar{u}}{\omega_b^2}.
\]
By substituting this solution ansatz into the stationary Fokker-Planck equation, we find a simple second order differential equation for $\zeta$ in terms of $\xi$,
\begin{equation}
\frac{\partial^2 \zeta}{\partial \xi^2} + \Lambda_{\pm} \xi \frac{\partial \zeta}{\partial \xi} = 0,
\end{equation}
where 
\begin{equation}
\Lambda_{\pm} = \frac{(\alpha - a_{\pm})\omega_b^2}{(1+a_\pm^2)\alpha \tilde{T}}
\end{equation}
and $a_\pm$ correspond to the roots of the quadratic equation $1+ \alpha a_\pm = a_\pm\omega_b^2 (a_\pm-\alpha)$,
\begin{equation}
a_\pm = \frac{\alpha (1+\omega_b^2)}{2 \omega_b^2} \pm \frac{1}{2\omega_b^2} \sqrt{\alpha^2 (1+\omega_b^2)^2 + 4 \omega_b^2}.
\end{equation}
The physically correct solution that satisfies the appropriate boundary conditions for $\zeta$ requires $a_-$, giving
\begin{equation}
\zeta(\xi) = \frac{1}{2} \erfc \left(\sqrt{\Lambda_-}\xi \right),
\end{equation}
where $\erfc(x)$ is the complimentary error function, $\erfc(x) = (2/\sqrt{\pi}) \int_{x}^{\infty} dt \; e^{-t^2}$.

The probability current density describing the flow in the spatial coordinate $x$ is defined in terms of the drift and diffusion coefficients as~\cite{Risken:1989}
\begin{equation}
J_x = D_x P - \frac{\partial}{\partial x_j} D_{x,j} P.
\end{equation}
By inserting the stationary Fokker-Planck solution subject to the boundary conditions defined by the function $\zeta(\xi)$, we find the overall particle current to be
\begin{eqnarray}
j_b &=& \int_{-\infty}^{\infty} dv \; (v + \bar{u}) P_0 (x_b,v) \notag \\
	&& + \alpha \bar{T} \sqrt{\frac{\Lambda_-}{\pi}}  \int_{-\infty}^{\infty} dv \; \frac{P_0 (x_b,v)}{\zeta(\xi) } e^{-\Lambda_- \xi^2}.
\end{eqnarray}
By using the dummy integration variable $z=v+\bar{u}$ and the identities
\begin{eqnarray}
\int_{-\infty}^{\infty} dz \; z \, e^{-a z^2} \erfc(b z + c)  = - \frac{b \, e^{-a c^2/(a + b^2)} }{a \sqrt{a + b^2}}, \notag \\
\int_{-\infty}^{\infty} dz \; e^{-a z^2} e^{-b(z + c)^2}  = \sqrt{\frac{\pi}{a+b}} \, e^{-a b c^2/(a+b)}, \notag
\end{eqnarray}
we find an analytic expression for the particle current at $x_b$,
\begin{eqnarray}
j_b &=& \frac{1}{2Z} \exp\left\{\frac{1}{\tilde{T}} \left[ -U(x_b) +  \bar{u} \beta x_b + \frac{\bar{u}^2}{2}\left(1 - \frac{\beta^2}{\omega_b^2}\right)   \right] \right\} \notag \\ 
&& \times \frac{1}{\omega_b} \left( \alpha \bar{T} \, \left[ 1-\Omega_b - \omega_b^2 (1+\Omega_b^2) \right]   \right),
\label{eq:jb}
\end{eqnarray}
where we have defined 
\begin{equation}
\Omega_b \equiv \sqrt{1 + \left[ \frac{2 \omega_b}{\alpha (1 + \omega_b^2)}  \right]^2 }
\end{equation}
to simplify the notation.

After combining (\ref{eq:na}) and (\ref{eq:jb}) with (\ref{eq:tau}), we recover an Arrhenius law for thermal activation over a single energy barrier,
\begin{equation}
\frac{1}{\tau} = \frac{1}{\tau_0} \exp \left[- \frac{\epsilon_{b}(\bar{u})}{\tilde{T}} \right],
\label{eq:Arrhenius}
\end{equation}
but with a current-dependent effective energy barrier
\begin{equation}
\epsilon_b(\bar{u}) = \epsilon_{b,0} - (\beta \Delta x) \bar{u} + \frac{1}{2}\left(\frac{1}{\omega_a^2} + \frac{1}{\omega_b^2} \right) (\beta\bar{u})^2
\label{eq:barrier_reduced}
\end{equation}
with linear and quadratic terms in the applied current (through $\bar{u}$), and an attempt frequency of the form
\begin{equation}
\frac{1}{\tau_0} = \frac{\omega_a}{2 \pi} \left( \frac{\alpha\; [1-\Omega_b - \omega_b^2 (\Omega_b -1)]}{2 \omega_b} \right).
\label{eq:tau0}
\end{equation}
Eqs.~\ref{eq:Arrhenius}-\ref{eq:tau0} constitute the main result of this paper. Because $\bar{u} \ll 1$ for typical low-current measurements in experiment, the dominant contribution to the current-dependence of the energy barrier is expected to be the linear term. It is important to note that both the linear and quadratic terms depend \emph{only} on the nonadiabatic coefficient $\beta$, which suggests that nonadiabatic torques can be quantified experimentally by measuring how the energy barrier varies with applied currents.

\section{Application to point defects}

In this section, we apply the theory developed in the preceding section to a simple model of domain wall pinning. Two kinds of point defects are considered: (1) A reduction in the uniaxial anisotropy. (2) A local hard axis pinning field. Both defects lead to a local minimum in the potential energy of the domain wall, which leads to pinning of the wall at the defect. From this simple model, it is possible to calculate directly important parameters for the Arrhenius transition rate, such as the field and current variation of the attempt frequency, barrier height, and barrier curvatures, which facilitates comparison of the present theory with experimental data. 

For the anisotropy defect, which is characterized by a pinning strength $H_p^{\rm ad}$ and assumed to be located at the origin, the corresponding energy is
\begin{eqnarray}
E_d^{\rm ad} &=&   - z_{\rm ad} \mu_0 H_p^{\rm ad} M_s \int dV \; \sin^2\theta[(X-X_0)/\lambda]\delta(X), \notag \\
&=& -z_{\rm ad} \mu_0 H_p^{\rm ad} M_s S_c \lambda \; \sech^2(x).
\end{eqnarray}
Despite the pointlike nature of the defect the potential well seen by the domain wall has a finite width because the wall has a finite spatial extension.~\cite{Gaunt:PMB:1983} For the field defect, which is characterized by $H_p^{\rm fd}$, a similar expression is obtained, 
\begin{eqnarray}
E_d^{\rm fd} &=&   - z_{\rm fd} \mu_0 H_p^{\rm fd} M_s \int dV \; \sin\theta[(X-X_0)/\lambda]\delta(X), \notag \\
&=& -z_{\rm fd} \mu_0 H_p^{\rm fd} M_s S_c \lambda \; \sech(x).
\end{eqnarray}
In the presence of an applied field $H_a$ along the positive $z$ axis, there is an additional Zeeman term in the potential energy of the form
\begin{equation}
E_z = -2 \mu_0 H_a M_s S_c X_0.
\end{equation}
The coefficient $z$ in each of the pinning energies is chosen such that the energy barrier vanishes for $H_a = H_p$. In terms of the scaled field $h \equiv H_a / H_p$, the total (scaled) potential energy $U = (1/E_0)(E_d + E_z)$ for the domain wall are
\begin{equation}
U_{\rm ad}(x) = - \frac{H_p^{\rm ad}}{M_s}\left( \frac{3\sqrt{3}}{4} \sech^2(x) + h x \right)
\end{equation}
and
\begin{equation}
U_{\rm fd}(x) = - \frac{H_p^{\rm fd}}{M_s}\left( 2 \sech(x) + h x \right).
\end{equation}
for the anisotropy and field defects, respectively.

\begin{figure}
\includegraphics[width=8.5cm]{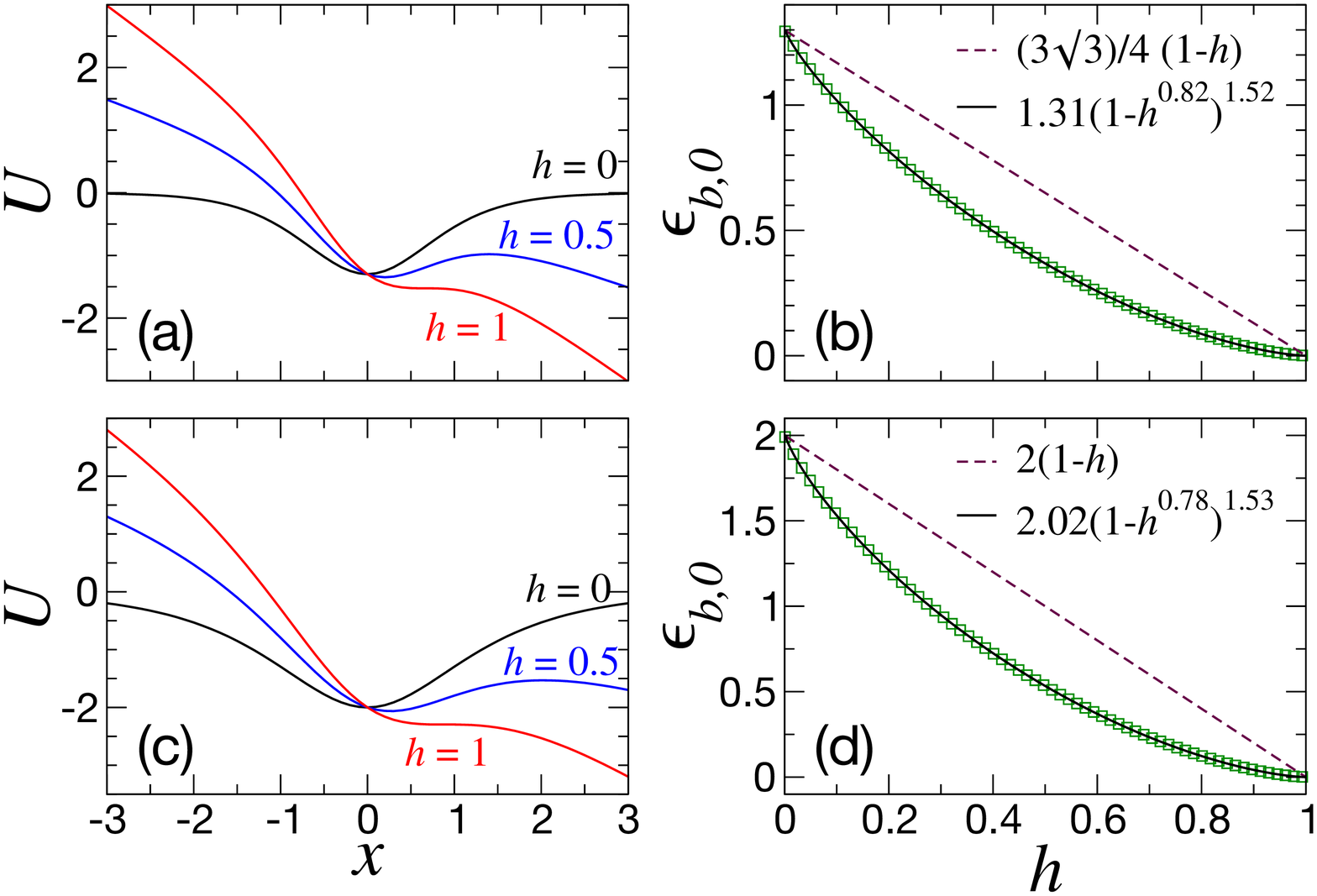}
\caption{\label{fig:energies}(Color online) (a) Domain-wall potential energy $U$ for three different applied fields with $H_p/M_s = 1$. (b) Field dependence of the energy barrier $\epsilon_{b,0}$, where squares correspond to the numerically calculated values and the solid line represents a fit of the form $\epsilon_{b,0} = C (1-h^\mu)^\nu$.}
\end{figure}

The potential profile for both anisotropy and field defects is shown in Fig.~\ref{fig:energies}a and \ref{fig:energies}c, respectively, as the applied field is varied from zero to a magnitude equal to the pinning field at which the barrier disappears.
%
It is straightforward to calculate the field dependence of the energy barrier with these pinning potentials. These results are shown in Fig.~\ref{fig:energies}b and \ref{fig:energies}d. The field variation of the energy barrier can be described empirically by
\begin{equation}
\epsilon_{b,0} = C (1-h^{\mu})^\nu,
\end{equation}
where $\mu,\nu$ are exponents. From fits using this equation, we note that $\nu \simeq 3/2$, which corresponds to a strong pinning form.~\cite{Gaunt:JAP:1986} For each defect, we have also included the simple linear field variation for the energy barrier~\cite{Attane:PRL:2006} for the purposes of comparison. While the linear form captures the general trends qualitatively, there is a large quantitative discrepancy between the exact form of $\epsilon_{b,0}(H)$ and the linear approximation, at least for the point defects considered here. Point defects therefore constitute an important class of pinning potentials for which a linear field/current equivalence does not apply for nonadiabatic torques.

\begin{figure}
\includegraphics[width=8.5cm]{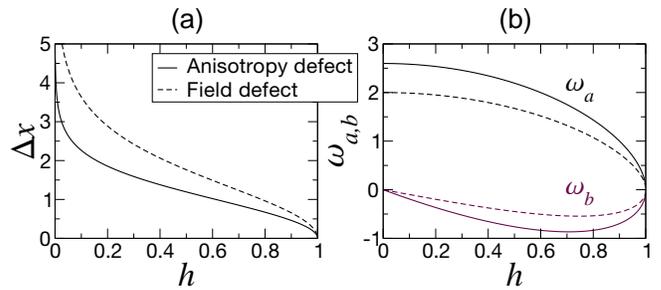}
\caption{\label{fig:delx_omegas}(Color online) Field depedence of (a) barrier width $\Delta x$ and (b) characteristic frequencies $\omega_{a,b}$. The latter are proportional to the curvature of the potential well and energy barrier, respectively. Solid lines represent the anisotropy defect, dashed lines represent the field defect.}
\end{figure}

Using this model, it is also possible to easily determine the field variation of the width of the potential well and the curvature of the potential well and barrier summit, given by the frequencies $\omega_a$ and $\omega_b$, respectively. These are presented in Fig.~\ref{fig:delx_omegas}.
%
The width of the barrier is of the same order of magnitude as the wall width $\lambda$, as expected, with a weak dependence on $h$ over a large range of applied fields. Because the linear current term in the energy barrier is also proportional to $\Delta x$, it may be important to account for such field variations in comparisons with experimental data. Nevertheless, one can still obtain a good order-of-magnitude estimate of nonadiabatic effects by assuming $\Delta x \sim 1$.

\begin{figure}
\includegraphics[width=8.5cm]{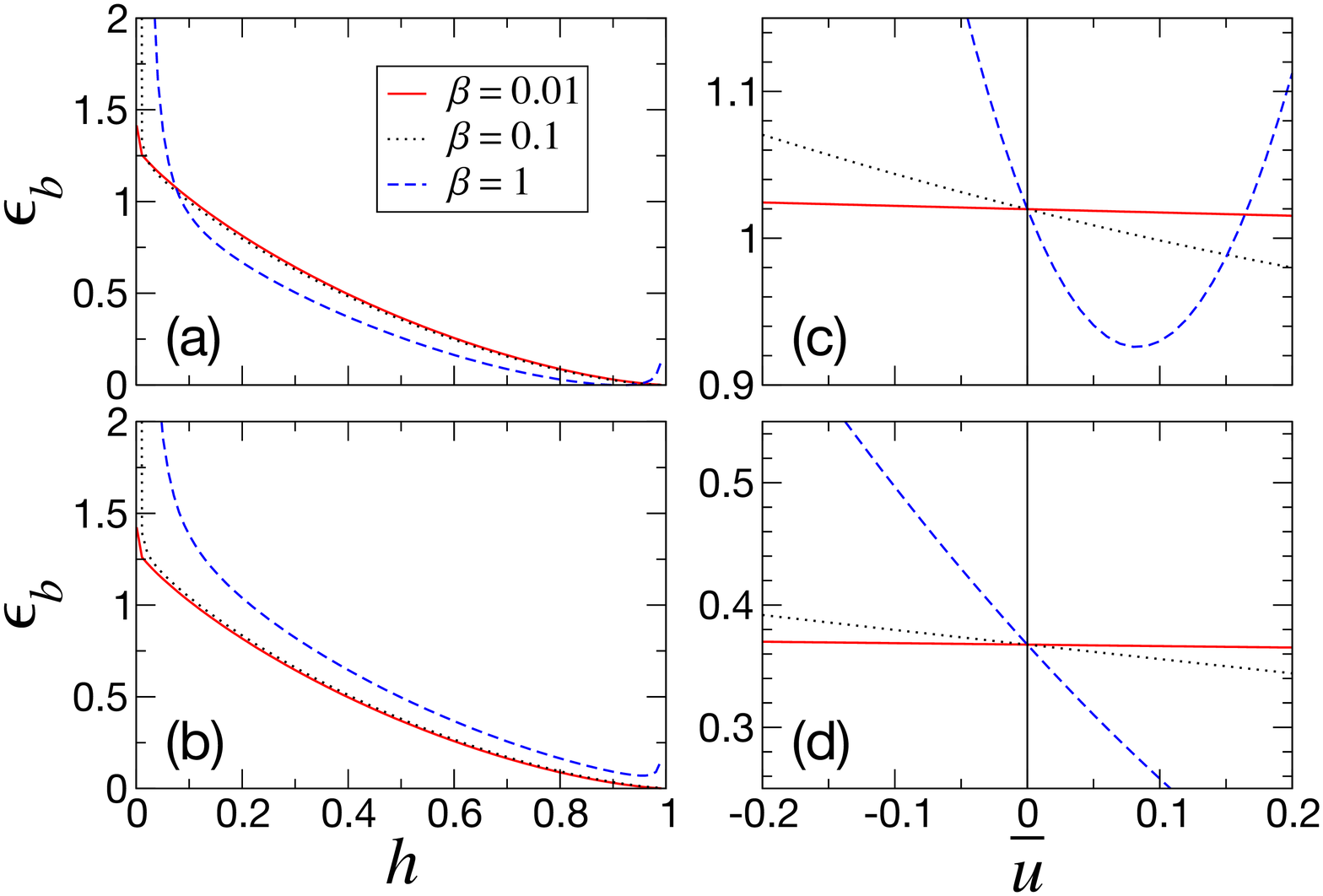}
\caption{\label{fig:barrier} (Color online) Energy barrier $\epsilon_b$ (in reduced units) as a function of (a,b) applied field $h$, with (a) $\bar{u} = 0.1$ and (b) $\bar{u} = -0.1$, and (c,d) applied current $\bar{u}$, with (c) $h=0.1$ and (d) $h=0.5$, for three values of the nonadiabatic torque coefficient $\beta$.}
\end{figure}

Some of the key features of the influence of spin torques on the energy barrier (\ref{eq:barrier_reduced}) are presented in Fig.~\ref{fig:barrier} for an anisotropy defect.
%
In Fig.~\ref{fig:barrier}a and \ref{fig:barrier}b, the field variation of the energy barrier is shown for positive and negative currents, respectively, for different values of the nonadiabatic torque coefficient. While the variations to the zero current limit are relatively small over most of the field range, there is a significant contribution from the quadratic current term near $h=0$ and $h=1$ at which the curvature of the barrier summit $\omega_b$ tends toward zero. This behavior is particularly explicit for the large $\beta = 1$ case considered. We reiterate that the present theory is only valid regions in which a well-defined potential well exists, i.e. for $\omega_{a,b} \neq 0$; the divergences at $h=0$ and $h=1$ are therefore unphysical. Nevertheless, we do expect that the quadratic current term to become important, particularly for large $\beta$ values, both at small fields and at fields close to the depinning field. This can be seen more explicitly in Figs.~\ref{fig:barrier}c and d in which the current dependence of the energy barrier is shown at weak and intermediate fields, respectively.

\section{Discussion and concluding remarks}

We have found that the transition rate associated with domain-wall depinning transitions driven by thermal activation, under applied magnetic fields and spin-polarized currents, can be described by an Arrhenius law in which the effective energy barrier depends on applied currents $I$ in the following way,
\begin{equation}
E_b(I) = E_{b,0} - \sigma_1 I + \sigma_2 I^2,
\end{equation}
where the two efficiency parameters $\sigma_{1,2}$ are defined as
\begin{subequations}
\begin{eqnarray}
\sigma_1 &=& \beta P \frac{\hbar}{e} \frac{\Delta X}{\lambda}, \\
\sigma_2 &=& \frac{1}{E_0} \left( \beta P \frac{\hbar}{e} \right)^2 \left(\frac{1}{\omega_a^2} +  \frac{1}{\omega_b^2}  \right).
\end{eqnarray}
\end{subequations}
Under usual experimental conditions the dominant contribution is expected to come from the linear term, which is proportional to the nonadiabatic coefficient $\beta$ and the spin-polarization of the material. The linear term also depends on the ratio between the width of the potential well and the domain wall width, $\Delta X/\lambda$, but this ratio is expected to be of the order of unity for pointlike defects. The quadratic current contribution can become important for small applied fields and for fields close to the switching field, where the curvature at the energy barrier summit ($\omega_b^2$) becomes small. However, we expect that it would be unrealistic to measure the statistics of depinning under vanishing magnetic fields (i.e., large energy barriers) as the transition rates under these conditions would be immeasurable in practice. In comparison with the effective temperature model proposed for thermally-assisted magnetization reversal with spin-transfer torques in magnetic nanopillars,~\cite{Li:PRB:2004} we note that spin-torque effects here appear also to renormalize the energy barrier (or the effective temperature), but with a different functional form that cannot be expressed in terms of a critical current $(1- I/I_c)$.

The transition rate has been obtained by assuming low domain-wall velocities for which we have assumed that the tilt angle $\phi$ is proportional to a streaming wall velocity $v$. This approximation is valid for wall propagation below Walker breakdown;~\cite{Schryer:JAP:1974} above the Walker threshold, wall motion is oscillatory by which the angle $\phi$ undergoes a precessional motion. In metallic alloys with perpendicular anisotropy, it is possible that pinning fields exceed the Walker field,~\cite{Metaxas:PRL:2007} which means that wall propagation would immediately proceed in the Walker regime following depinning from a defect. We believe that the present theory remains valid even for large pinning fields exceeding the Walker field, because the motion associated with the depinning transition is more likely to be viscous rather than precessional. This can be understood as follows. The Walker regime is attained when the slope of the domain wall potential $-\partial U/\partial x$, which corresponds to a linear force on the wall, exceeds a certain value. Because the slope of the potential well decreases to zero at the pinning center, any domain wall motion driven by thermal fluctuations around this position is likely to result in viscous motion rather than precessional motion because the net force acting on the wall remains small. This is equally true at the barrier summit. Because the transition rate is ultimately determined by the dynamics at the bottom of the potential well and at the barrier summit, we expect the approximation of viscous motion will remain relatively robust in spite of possible thermally-driven excursions into the Walker regime.

We wish to reiterate that the current-induced modifications to the energy barrier do not appear necessarily as an effective field. This is not entirely surprising given that the nonadiabatic torques enter only as a fieldlike contribution in the equations of motion, but cannot themselves be derived from a potential because they are nonconservative torques. As such, we contend that measures of the $\beta$ coefficient through fieldlike variations, e.g. associating a change in propagation field with a nonadiabatic torque, can lead to erroneous estimates if thermal activation or magnetic viscosity is not accounted for properly. This problem is particularly important for ferromagnetic alloys with perpendicular anisotropy where we expect the influence of magnetic viscosity, due to the large contribution of intrinsic defects, to be important for current-driven wall dynamics. We suggest that measurements of the current-dependence transition rate of domain-wall depinning from defects should provide a more direct and accurate determination of the nonadiabatic torque in experiment.

\begin{acknowledgments}
We thank D. Ravelosona and C. Chappert for stimulating discussions. This work was partially supported by the ISTRADE contract of the French National Research Agency (ANR). C. B. acknowledges financial support from the regional government of {\^I}le-de-France through C'Nano IdF.
\end{acknowledgments}

\bibliography{articles}

\end{document}